\documentclass[aps,prl,preprint,groupedaddress]{revtex4-2}

\usepackage{physics}
\usepackage{amssymb}
\usepackage[dvipdfmx]{graphicx}
\usepackage{here}
\usepackage{MnSymbol}

\begin{document}


\title{Quantum-inspired information entropy in multi-field turbulence}


\author{Go Yatomi}
\email{yatomi.go@nifs.ac.jp}
\affiliation{SOKENDAI (The Graduate University for Advanced Science), Gifu, 509-5292, Japan}
\author{Motoki Nakata}
\affiliation{Faculty of Arts and Sciences, Komazawa University, Tokyo 154-8525, Japan \\
RIKEN Interdisciplinary Theoretical and Mathematical Sciences Program (iTHEMS), Saitama 351-0198, Japan \\
PRESTO, Japan Science and Technology Agency, Saitama 332-0012, Japan}


\date{\today}

\begin{abstract}
  A novel information entropy of turbulence systems with multiple field quantities is formulated. Inspired by quantum mechanics, the von Neumann entropy (vNE) and the entanglement entropy (EE) are derived from a density matrix for the turbulence state in terms of the multi-field singular value decomposition (MFSVD). Applying the information-theoretic entropy analyses to spatio-temporal dynamics in turbulent plasmas with phase-transition-like behavior, we discover a new nontrivial transition threshold regarding the vNE, which significantly deviates from the transition threshold of the field energy considered in the conventional approaches. These findings provide us with new classifications of the turbulence state in terms of combined energy and information. It is also shown that the EE for nonlinear interactions in turbulence simultaneously describes not only the information for the strength of nonlinear mode couplings but also the direction of net energy transfer.
\end{abstract}


\maketitle

\textit{Introduction}.---''Information'' from a probabilistic point of view can provide us with new physical insights to extract essential properties in complicated dynamics of fields and particles. In these decades, information theory has been extensively combined into physical analyses in, e.g., quantum mechanics \cite{Vedral1997,Horodecki2009}, the quantum gravity theory of black holes \cite{Bekenstein1973,Belokolos2009}, the non-equilibrium thermodynamics \cite{Bennett1987,Sagawa2008}, and the statistical mechanics of critical phenomena in many-body systems \cite{Holzhey1994,Dong2008,Matsueda2012}. \par
Not just in quantum mechanics, information-theoretic entropy analyses are also utilized for the studies of turbulent fluids and plasmas. Falkovich \textit{et al.} introduced a relative entropy for homogeneous steady turbulence in shell-averaged wavenumber space, based on multi-mode correlations with the Gaussian statistics \cite{Falkovich2023}. Tanogami and Araki explored the nature of information flow in turbulent energy cascade processes, where the time derivative of the mutual information for the joint probability density function of the turbulence intensity at neighboring shell-averaged wavenumbers is formulated \cite{Tanogami2024}. Also, the von Neumann entropy for the time series data observed in laboratory plasma turbulence has been investigated  \cite{Kawamori2017}. 
It is noted that the previous works examined the information entropy with respect to the integrated turbulence intensity and the temporal evolution. Thus, the inhomogeneous and/or anisotropic nature of spatial patterns of the turbulence is ignored. Extensions towards the simultaneous treatment of multiple components in turbulent fields are also yet achieved. \par
In this paper, we propose a novel information-theoretic entropy analysis for more general turbulent fields, by constructing a density matrix with a unitary basis reflecting spatio-temporal structures of multiple turbulent fields and coherent flow patterns. The formulation is based on our previous work of constructing ''multi-field'' singular value decomposition (MFSVD), which is suitable for decomposing the multi-component inhomogeneous and/or anisotropic fields \cite{Yatomi2023,Kodahara2023}. The density matrix of turbulence enables us to derive straightforwardly the von Neumann entropy (vNE) and the entanglement entropy (EE), the quantity for the coupling of subsystems, on the analogies of quantum mechanics and quantum information theory. As will be shown later, these entropy analyses discovered a new nontrivial classification/description of the phase transition and nonlinear correlations in multi-component turbulence. \par
As a representative model of a multi-component turbulence system with transitional behavior, we consider the Hasegawa-Wakatani equations \cite{Hasegawa1983,Numata2007}. The nonlinear advection in the model can produce the transition between incoherent vortices and coherent sheared-flow patterns, so-called zonal flows (ZF). The Hasegawa-Wakatani plasma turbulence is, thus, a good measure to examine our new formulations on vNE and EE of turbulence. \par

\textit{Information entropies of turbulence}.---How to construct the density matrix and associated vNE and EE for multi-field turbulence is given as follows. Here, we consider two-dimensional turbulence fields (the extension to the higher dimensional case is straightforward). Let $f_k(x,y,t)$ ($k=1,\cdots,M$) be the $k$-th field component, e.g., $k=1$ for the particle density field, $k=2$ for the velocity field, etc. After discretizing on the spatial grids, the $k$-th field at time $t$ is rearranged to a one-dimensional vector $\vb{f}_k(t) = (f_k(x_1,y_1,t),\cdots,f_k(x_{N_x},y_1,t),\cdots,f_k(x_1,y_{N_y},t),\cdots,$\\$f_k(x_{N_x},y_{N_y},t))^\mathrm{T}$, where $N_x$ and $N_y$ are the grid numbers in the $x$ and $y$ direction, respectively. Then, the time series of vectors for $k=1,\cdots,M$ are combined into a matrix $F$; $F = (\vb{f}_1(t_1),\cdots,\vb{f}_M(t_1),\cdots,\vb{f}_1(t_{N_t}),\cdots,\vb{f}_M(t_{N_t}))$, where $N_t$ is the number of discretized time points. Note that the matrix $F$ simultaneously contains the spatio-temporal data for the multiple turbulence fields of interest. By applying SVD, the matrix $F$ is decomposed by a diagonal matrix of singular values $\Sigma=\mathrm{diag}(s_1,\cdots,s_N)$ and two unitary matrices $U$ and $V$, i.e., $F=U\Sigma V^\dagger$, where $N=\min(N_xN_y, MN_t)$ is the number of SVD modes. Inversely converting from the matrix form to the continuous field representation, the original $k$-th field quantity $f_k(x,y,t)$ is decomposed by means of the orthonormal spatial basis $\psi_i$ and the coefficient of the temporal evolution $h_i^{(k)}$:
\begin{equation}
  f_k(x,y,t) = \sum_{i=1}^{N} s_ih_i^{(k)}(t)\psi_i(x,y),
\end{equation}
where $h_i^{(k)}$ depends on the field index $k$. We emphasize that $s_i$ and $\psi_i$ are common to all fields, and thus $\psi_i$ reflects the multi-field correlation, preserving the orthonormality of the spatial structures:
\begin{equation}
  \int\dd\vb{x}\psi_i\psi_j=\delta_{ij}.
\end{equation}
This is the basic outline of MFSVD, and one can find more details in Ref. \cite{Yatomi2023}. \par
\begin{figure}[tb]
  \centering
  \includegraphics[width=8cm]{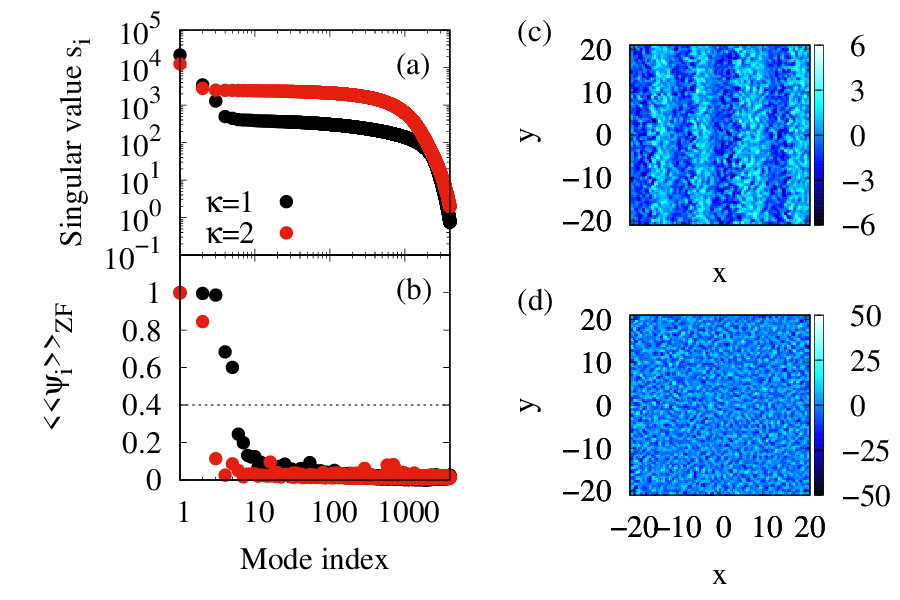}
  \caption{The simulation results of the Hasegawa-Wakatani equation and the multi-field SVD in the case of $(\kappa,\alpha) = (1,1)$ and $(2,1)$. (a) shows the SVD-spectrum of the singular value $s_i$, (b) shows the zonal-flow amplitude $\llangle\psi_i\rrangle_\mathrm{ZF}$. (c) and (d) are the snapshots of $\zeta$ in the case of the zonal-flow-dominated state $(\kappa,\alpha) = (1,1)$ and the turbulence-dominated state $(2,1)$, respectively.}
  \label{fig:info}
\end{figure}

Since the orthonormal spatial structure $\psi_i(x,y)$, which is resulting from the unitary matrix $U$, can be regarded as a basis in the Hilbert space of physical field quantities in $(N_xN_y)$ dimension. Hereafter, we denote this basis using bra-ket notation such as $\ket*{\psi_i}=\psi_i(x,y)$. Then, we can define a mixed state for a subsystem $X\subset\Omega=\{1,2,\cdots,N\}$ in the whole SVD mode space $\Omega$ as follows:
\begin{align}
  \rho_X &:= \sum_{i\in X} \eta_i \ket*{\psi_i}\bra*{\psi_i}, \\
  \eta_i &:= \left.(s_ih_i^{(k)})^2 \middle/ \sum_{i\in X}\qty(s_ih_i^{(k)})^2\right.,
\end{align}
where $\sum_{i\in X}\eta_i = 1$. Once we obtained the density matrix $\rho_X$, the von Neumann entropy (vNE) $S_\mathrm{vN}^X$ is derived in a  similar manner to  quantum systems:
\begin{equation}
  S_\mathrm{vN}^X := - \Tr(\rho_X \ln{\rho_X}) = - \sum_{i \in X} \eta_i \ln{\eta_i}. \label{eq:vNe}
\end{equation}
The typical spectrum of the singular value $s_i$ is shown in Fig. \ref{fig:info}(a) (explained in detail later). Since $h_i^{(k)}\sim O(1)$, the vNE is approximately regarded as the Shannon entropy reflecting the broadness of the SVD mode spectrum $s_i$ under the appropriate normalization. \par

The vNE is for a mixed state in the subsystem of multi-field turbulence. On the other hand, we can define a compound state in two subsystems $A,B\subset\Omega$ as a tensor product form:
\begin{align}
  \ket*{\psi_{AB}} &:= \sum_{i\in A}\sum_{j\in B} \sigma_{ij} \ket*{\psi_i} \otimes \ket*{\psi_j}, \label{eq:CompoundState}
\end{align}
where $\sigma_{ij}$ denotes the coefficients such that $\sum_{i,j}(\sigma_{ij})^2=1$. The entanglement entropy (EE) of the compound state $S_\mathrm{EE}$ is derived as the vNE of the reduced density matrix:
\begin{align}
  \rho_A &= \Tr_B \ket*{\psi_{AB}}\bra*{\psi_{AB}}, \\
  S_\mathrm{EE} &:= - \Tr_A (\rho_A \ln{\rho_A}). \label{eq:EE}
\end{align}
Note that, in the case of the subsystems being ''entangled'', the coefficients $\sigma_{ij}$ must satisfy the form of $\sigma_{ij}\neq a_ib_j$. If it is expressed as a simple product form for each subsystem, i.e., $\sigma_{ij} = a_ib_j$, then the compound state is called a ''separable'' state, where the EE is equal to zero. The proper form of $\sigma_{ij}$ is chosen to describe the nonlinear correlation of interest, as will be discussed in Eq.(\ref{eq:sigma_Skij}). \par

\textit{Hasegawa-Wakatani equations}.---The present entropy analyses are applied to a multi-component plasma turbulence, where the phase transition and the nonlinear correlation are prominent. The two-dimensional drift-wave turbulence with zonal-flow formations in magnetized plasmas are described by the (modified) Hasegawa-Wakatani equation in the normalized form: \cite{Hasegawa1983,Numata2007}
\begin{align}
  \partial_t \zeta&=\{\zeta,\phi\}+\alpha\qty(\tilde{\phi}-\tilde{n})-D\nabla^4\zeta, \label{eq:HW_zeta} \\
  \partial_t n&=\{n,\phi\}+\alpha\qty(\tilde{\phi}-\tilde{n})-\kappa\partial_y\phi - D\nabla^4 n, \\
  \nabla^2 \phi &= \zeta,
\end{align}
where $\nabla^2=\partial_x^2+\partial_y^2$, $\{A,B\}=\partial_xA\partial_yB-\partial_yA\partial_xB$. $\tilde{A}=A-\expval{A}$ is the non-zonal component of the quantity $A$, where $\expval{A}=1/L_y\int\dd yA$ is the zonal average. The multiple field quantities of the electrostatic potential $\phi$, the vorticity $\zeta$, and the electron density $n$ are considered. \par
The turbulence dynamics bifurcates, depending on three parameters; the density gradient $\kappa$ as the driving source of the linear instability, the adiabatic parameter $\alpha$ controlling the coupling strength of the two fields (associated with the electron motion along the magnetic field), and the dissipation coefficient $D$. The typical simulation results of the zonal-flow dominated and the turbulence-dominated quasi-steady states are shown in Fig. \ref{fig:info}. \par

Since the spontaneous zonal-flow formation occurs by the nonlinear interactions between $\phi$ and $\zeta$, MFSVD is applied to the simulation data for the quasi-steady state. Then we divide the SVD modes into two subsystems, i.e., the zonal modes and the non-zonal modes, according to a threshold of the relative zonal-flow amplitude of the mode $\llangle\psi_i\rrangle_\mathrm{ZF}$; $Z = \{ i\, |\llangle\psi_i\rrangle_\mathrm{ZF} > 0.4\}$ and $T = \{ i\, |\llangle\psi_i\rrangle_\mathrm{ZF} < 0.4\}$, where $\llangle\psi_i\rrangle_\mathrm{ZF} = \int \dd x \qty(\int \dd y\, \psi_i)^2 / \int dxdy\,(\psi_i)^2$. By doing this, one can examine vNE for the zonal flow and the ambient turbulence separately. \par

\begin{figure}[tb]
  \centering
  \includegraphics[width=8cm]{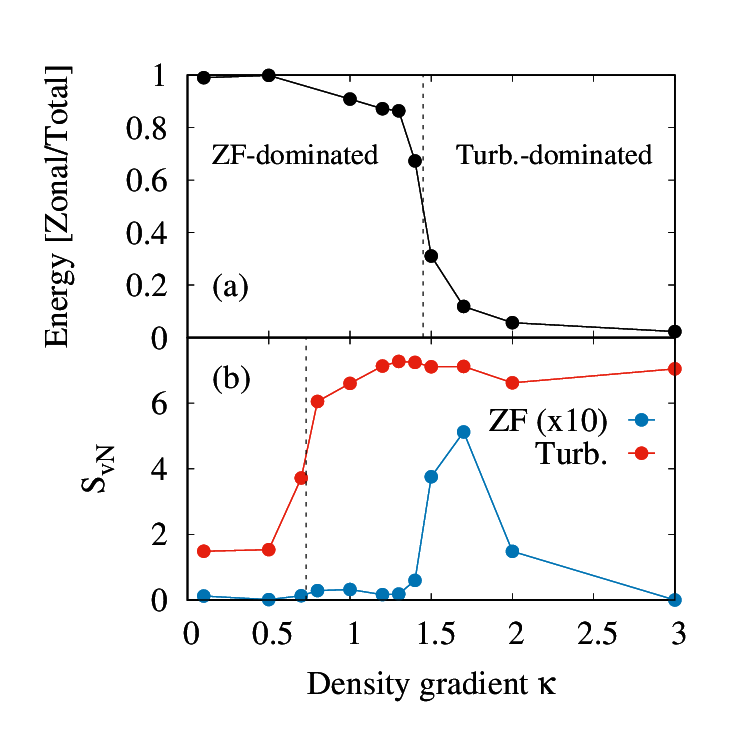}
  \caption{(a) $\kappa$ dependence of the relative zonal-flow energy ratio, and (b) the von Neumann entropy of the zonal-flow subsystem (blue) and the turbulence subsystem (red).}
  \label{fig:kappa-zfr-vne}
\end{figure}

\textit{A new transition threshold of von Neumann entropy}.---Figure \ref{fig:kappa-zfr-vne}(a) shows the $\kappa$ dependence of the relative energy ratio of the zonal flow, indicating a clear transition at $\kappa\simeq1.5$, defined by the point where $\mathrm{(ZF\,energy)/(total\,energy)}=0.5$. This is the conventional way of identifying the phase transition between the zonal-flow-dominated and turbulence-dominated states (e.g., Ref. \cite{Numata2007}). For fixed $\alpha=1$, the zonal flow is strongly generated in the region of $\kappa<1.5$ with the moderate linear growth rate of the turbulent fluctuations, while the zonal flow is suppressed in the large $\kappa$ region where the ambient turbulence strongly grows. \par
The $\kappa$ dependence of vNE for the turbulence subsystem $S_\mathrm{vN}^T$ and the zonal-flow subsystem $S_\mathrm{vN}^Z$ are shown in Fig. \ref{fig:kappa-zfr-vne}(b). $S_\mathrm{vN}^T$ is considerably larger than $S_\mathrm{vN}^Z$ in the whole region of $\kappa$. This is because the zonal-flow subsystem consists of only several SVD modes as shown in Fig. \ref{fig:info}(a), while the turbulence subsystem is composed of a number of SVD modes with a broader distribution. In contrast to the conventional energy-ratio argument, another kind of the phase transition with respect to vNE is discovered in the turbulence subsystem, where the transition threshold, $\kappa \simeq 0.7$, is clearly smaller than that in the energy ratio, $\kappa \simeq 1.5$. Here, the threshold of $S_\mathrm{vN}^T$ is defined by the $\kappa$ value where the $S_\mathrm{vN}^T$ bisects the maximun and the minimum values with $\alpha$ fixed. It is also noted that $S_\mathrm{vN}^Z$ well captures the energy-ratio transitions as well. Indeed, for $\kappa < 1.5$, nearly zero value of $S_\mathrm{vN}^Z$ is associated with a significantly shrunk SVD mode spectrum. Around the critical value in the energy ratio, $\kappa\simeq1.5$, $S_\mathrm{vN}^Z$ increases because of the broadening of the SVD mode spectrum, which is physically interpreted as a perturbed state with the various scales of zonal flows. In the large $\kappa$ limit where $S_\mathrm{vN}^Z$ decays, the number of the zonal-flow modes quickly decreases. \par
\begin{figure}[tb]
  \centering
  \includegraphics[width=9cm]{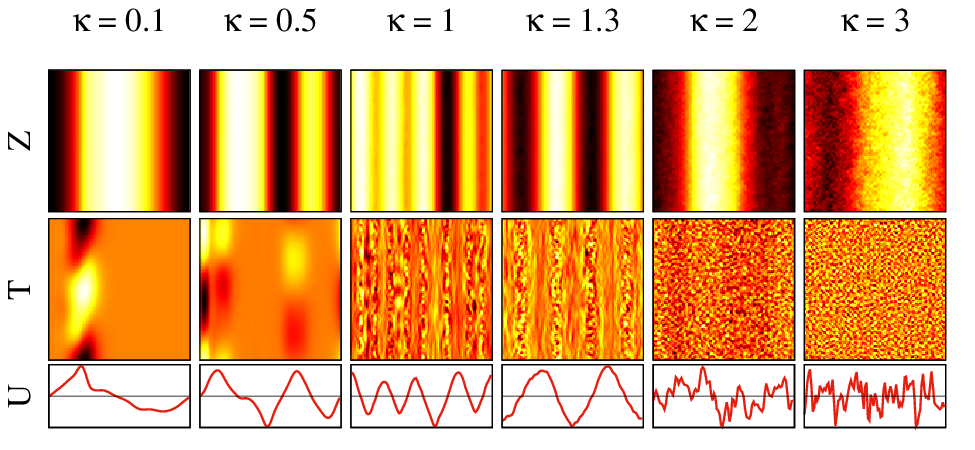}
  \caption{The contour of the dominant pattern of each subsystem $\phi_T(x,y)$ and $\phi_Z(x,y)$ and the zonal-flow velocity $U$ (with black lines representing $U=0$) for several $\kappa$ values.} 
  \label{fig:psi2x4}
\end{figure}

Based on these findings from quantum-inspired information entropy analyses, we can identify new classification of states in plasma turbulence, i.e., [State i] zonal-flow-dominated state with small $S_\mathrm{vN}^T$, [State ii] zonal-flow dominated state with large $S_\mathrm{vN}^T$, [State iii] transitional state with large $S_\mathrm{vN}^Z$, and [State iv] turbulence-dominated state with large $S_\mathrm{vN}^T$. The typical spatial structures for each case will be shown in Fig. \ref{fig:psi2x4}.

\begin{figure}[tb]
  \centering
  \includegraphics[width=8cm]{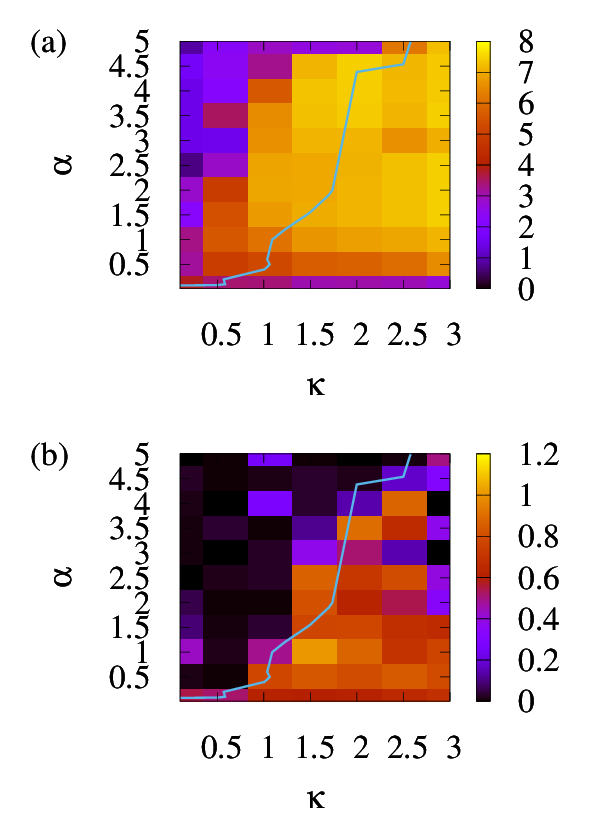}
  \caption{The coutours of the von Neumann entropy of (a) the turbulence subsystem and (b) the zonal-flow subsystem in the $\kappa$-$\alpha$ space. The cyan curve shows the boundary of the relative energy ratio where $\mathrm{(ZF\,energy)/(total\,energy)}=0.5$.}
  \label{fig:vnecounter}
\end{figure}

Figure \ref{fig:vnecounter} shows the contour plots of the vNE for turbulence and zonal-flow subsystems in the $\kappa$-$\alpha$ parameter space. The cyan curves represent the transition threshold in the conventional energy ratio, where the critical value of $\mathrm{(ZF\,energy)/(total\,energy)}=0.5$ is considered. The entropy transition boundary for $S_\mathrm{vN}^T$ clearly deviates from the energy transition boundary. Thus, it is repeatedly noted that there exists a novel classification of the quasi-steady states from the information point of view. On the other hand, the entropy transition boundary of $S_\mathrm{vN}^Z$ shown in Fig. \ref{fig:vnecounter}(b) reasonably coincides with the energy transition boundary, as explained in Fig. \ref{fig:kappa-zfr-vne}(b). \par 

The contour of the dominant pattern of each subsystem $\phi_Z(x,y)=\sum_{i\in Z}s_ih_i^{(\phi)}\psi_i(x,y)$ and $\phi_T(x,y)=\sum_{i\in T}s_ih_i^{(\phi)}\psi_i(x,y)$ is shown in Fig. \ref{fig:psi2x4}. The zonal-flow velocity profile $U(x)=\partial_x\expval{\phi}$ is also displayed. The two panels on the left correspond to [State i], the two panels in the middle correspond to [State ii], and the two panels on the right correspond to [State iv] as discussed in Fig. \ref{fig:kappa-zfr-vne}. 
For the turbulence subsystem, the vortex streets trapped in the crest of the zonal flow \cite{Nakata2010,Nakata2011} in [State i]. Because the large-scale vortices dominate the ambient turbulence, $S_\mathrm{vN}^T$ of these states is relatively small. On the other hand, the trapping effect of the zonal flow is lost as $\kappa$ increases, and then the relatively more homogeneous and small-scale fluctuations are generated, resulting in increases in $S_\mathrm{vN}^T$. This transition within non-zonal turbulent fluctuations occurs even in the zonal-flow dominated state. Although the trapping of turbulent vortices by the zonal flow is recognized in previous work \cite{Sasaki2018}, the bifurcation of the trapped and the un-trapped states is newly revealed quantitatively by the present information entropy analyses.

\textit{Entanglement entropy of nonlinear interaction}.---In both the Fourier wavenumber space and SVD mode space, the nonlinear interactions in the turbulence system are described by the triad energy transfer $\mathcal{J}_k^{i,j}$ \cite{Nakata2012,Yatomi2023}:
\begin{align}
  \mathcal{J}_k^{i,j} &:= \frac{1}{2} \int \dd\vb{x}\,\phi_k(\{\phi_i,\zeta_j\}+\{\phi_j,\zeta_i\}) \notag \\
  &= \frac{1}{2} s_is_js_kh_k^{(\phi)}\qty(h_i^{(\phi)}h_j^{(\zeta)}-h_j^{(\phi)}h_i^{(\zeta)})\int\dd\vb{x}\,\psi_k\{\psi_i,\psi_j\}.
\end{align}
Here, $(i,j,k)$ means the SVD mode indices. Note that $\mathcal{J}_k^{i,j}=\mathcal{J}_k^{j,i}$, and $\mathcal{J}_k^{i,j}+\mathcal{J}_j^{k,i}+\mathcal{J}_i^{(j,k)}=0$. \par
Now, we define the EE for this $\mathcal{J}_k^{i,j}$. Since the triad energy transfer $\mathcal{J}_k^{i,j}$ indicates the negative and positive values depending on the direction of energy flow, the exponential form is appropriate to the condition of $\sum_{i,j}(\sigma_{ij})^2=1$:
\begin{align}
  \sigma_{ij}^\pm := C\exp(\pm \mathcal{J}_k^{i,j}), \label{eq:sigma_Skij}
\end{align} 
where $C=\qty{\sum_{i,j}\exp(\pm \mathcal{J}_k^{i,j})}^{-\frac{1}{2}}$, and $k$ is arbitray but fixed here. Two expressions of the EE are then derived from Eq. (\ref{eq:EE}); $S_{EE}^+$ for $\sigma_{ij}^+$ and $S_{EE}^-$ for $\sigma_{ij}^-$. We choose this form of the coefficients in order not to make the state separable and to distinguish between inflows ($S_{EE}^+$) and outflows ($S_{EE}^-$) of the energy transfer. In this formulation, the EE is just a real scalar value that extracts the nonlinear coupling of turbulence fields as the information quantity. The EE enables the dimensionality reduction of the original $\mathcal{J}_k^{i,j}$, still retaining the information on the direction of the transfer and the structures in the mode space, which is lost by the simple summation by the indices (or wavenumbers). In similar to quantum mechanics, a large value of the EE means that only several limited modes are strongly coupled to drive energy transfer via the zonal mode. \par
\begin{figure}[tb]
  \centering
  \includegraphics[width=7cm]{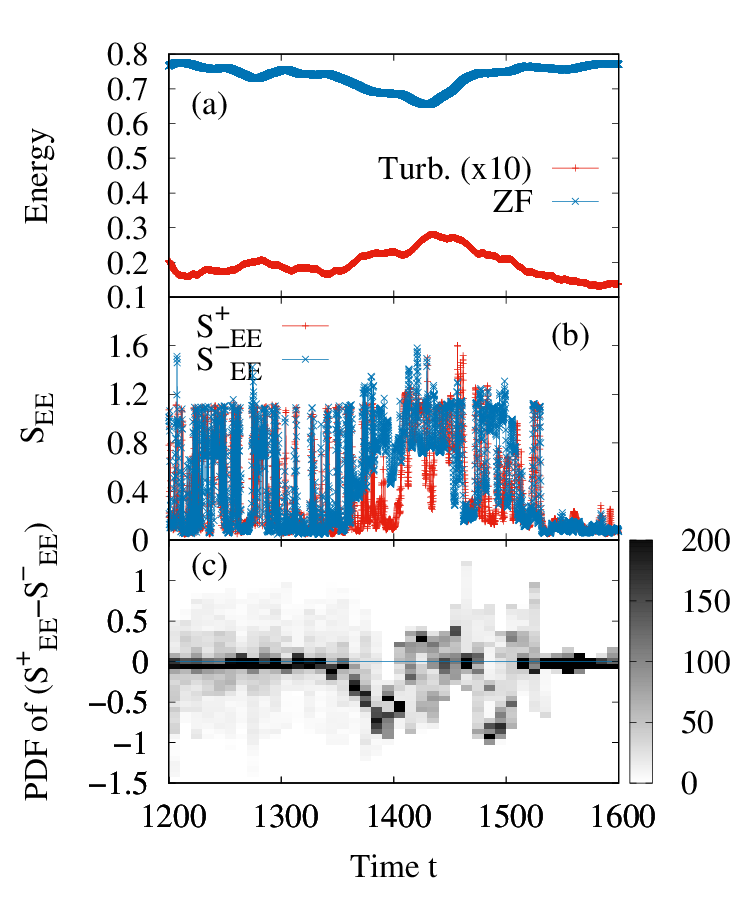}
  \caption{The temporal evolution of (a) the field energy of the turbulence (red) and the zonal flow (blue), (b) the entanglement entropy $S_\mathrm{EE}^+$ (red) and $S_\mathrm{EE}^-$, and (c) the probability density function of the difference of the entanglement entropy $S_\mathrm{EE}^+-S_\mathrm{EE}^-$ in the case of $(\kappa,\alpha)=(1,3)$.}
  \label{fig:ee_k1a3}
\end{figure}

Figure \ref{fig:ee_k1a3} shows the comparison of the temporal evolution of the field energy and the EE in the case of $(\kappa,\alpha)=(1,3)$. The mode $k$ is fixed to the zonal-flow mode $k=1$, so that we look into the nonlinear energy transfer mediated by the zonal flow. As shown in the panel (a), the field energy of the zonal flow changes in the slow time scale. However, the EE shown in the panel (b) oscillates in a much smaller time scale compared to the energy variation. Then, it is found that the slow variation of the field energy results from the accumulation of many instantaneous nonlinear interactions. \par
While both expressions of the EE $S_\mathrm{EE}^+$ and $S_\mathrm{EE}^-$ have similar values during periods of increasing and decreasing energy, the difference between them is shown in Fig. \ref{fig:ee_k1a3} as the temporal evolution of the probability density function (PDF) of $S_\mathrm{EE}^+ - S_\mathrm{EE}^-$. During the field energy decaying $(1350<t<1400)$, $S_\mathrm{EE}^-$ is relatively larger than $S_\mathrm{EE}^+$, and the PDF peaks in the negative region. Then, $S_\mathrm{EE}^+$ becomes dominant over $S_\mathrm{EE}^-$ when the field energy is growing $(1400<t<1470)$. When the growth rate of the ZF energy settles down at $t\simeq1480$, the PDF tends to be negative again, implying the saturation of growth. 
Therefore, the EE for the turbulence system provides the information of not only the strength of the nonlinear coupling but also the direction of the energy transfer. In other words, the present EE is a scalar quantity representing the information flow in the nonlinear interaction, but keeping the spatial structures in the unitary basis $\psi(x,y)$. Note that similar analyses to examine directly $\mathcal{J}_k^{i,j}$ including their temporal dynamics are often difficult in the conventional mode-by-mode approach. \par

\textit{Summary}.---In this paper, we formulate the quantum-inspired information entropy of multi-field turbulence. Using the multi-field singular value decomposition (MFSVD), the density matrix for the turbulence state is constructed. Then, the von Neumann entropy (vNE) and the entanglement entropy (EE) are derived. As a result, we discover a novel transition of the turbulence structure based on the information entropy, which cannot be seen by the conventional energy argument. Furthermore, the present EE can provide new insights on nonlinear interactions in turbulence. It should be emphasized that the proposed formulation can be applied to any other kinds of turbulent fields and nonlinear correlations, beyond the present turbulent plasmas. We also expect more extended applications of the VNE and the EE as a novel turbulence measurement principle, which will be addressed in future works.

We thank Makoto Sasaki and Masayuki Yokoyama for the helpful discussions. Numerical computations are performed on the NIFS Plasma Simulator. This work was supported in part by JST, the establishment of university fellowships towards the creation of science technology innovation, Grant Number JPMJFS2136, in part by JST, PRESTO Grant Number JPMJPR21O7, in part by the NIFS collaborative Research programs (NIFS23KIST039, NIFS23KIST044), and in part by PLADyS, JSPS Core-to-Core Program.

\bibliography{EntanglementEntropy}

\begin{thebibliography}{20}%
\makeatletter
\providecommand \@ifxundefined [1]{%
 \@ifx{#1\undefined}
}%
\providecommand \@ifnum [1]{%
 \ifnum #1\expandafter \@firstoftwo
 \else \expandafter \@secondoftwo
 \fi
}%
\providecommand \@ifx [1]{%
 \ifx #1\expandafter \@firstoftwo
 \else \expandafter \@secondoftwo
 \fi
}%
\providecommand \natexlab [1]{#1}%
\providecommand \enquote  [1]{``#1''}%
\providecommand \bibnamefont  [1]{#1}%
\providecommand \bibfnamefont [1]{#1}%
\providecommand \citenamefont [1]{#1}%
\providecommand \href@noop [0]{\@secondoftwo}%
\providecommand \href [0]{\begingroup \@sanitize@url \@href}%
\providecommand \@href[1]{\@@startlink{#1}\@@href}%
\providecommand \@@href[1]{\endgroup#1\@@endlink}%
\providecommand \@sanitize@url [0]{\catcode `\\12\catcode `\$12\catcode `\&12\catcode `\#12\catcode `\^12\catcode `\_12\catcode `\%12\relax}%
\providecommand \@@startlink[1]{}%
\providecommand \@@endlink[0]{}%
\providecommand \url  [0]{\begingroup\@sanitize@url \@url }%
\providecommand \@url [1]{\endgroup\@href {#1}{\urlprefix }}%
\providecommand \urlprefix  [0]{URL }%
\providecommand \Eprint [0]{\href }%
\providecommand \doibase [0]{https://doi.org/}%
\providecommand \selectlanguage [0]{\@gobble}%
\providecommand \bibinfo  [0]{\@secondoftwo}%
\providecommand \bibfield  [0]{\@secondoftwo}%
\providecommand \translation [1]{[#1]}%
\providecommand \BibitemOpen [0]{}%
\providecommand \bibitemStop [0]{}%
\providecommand \bibitemNoStop [0]{.\EOS\space}%
\providecommand \EOS [0]{\spacefactor3000\relax}%
\providecommand \BibitemShut  [1]{\csname bibitem#1\endcsname}%
\let\auto@bib@innerbib\@empty
\bibitem [{\citenamefont {Vedral}\ \emph {et~al.}(1997)\citenamefont {Vedral}, \citenamefont {Plenio}, \citenamefont {Rippin},\ and\ \citenamefont {Knight}}]{Vedral1997}%
  \BibitemOpen
  \bibfield  {author} {\bibinfo {author} {\bibfnamefont {V.}~\bibnamefont {Vedral}}, \bibinfo {author} {\bibfnamefont {M.~B.}\ \bibnamefont {Plenio}}, \bibinfo {author} {\bibfnamefont {M.~A.}\ \bibnamefont {Rippin}},\ and\ \bibinfo {author} {\bibfnamefont {P.~L.}\ \bibnamefont {Knight}},\ }\bibfield  {title} {\bibinfo {title} {{Quantifying Entanglement}},\ }\href {https://doi.org/10.1103/PhysRevLett.78.2275} {\bibfield  {journal} {\bibinfo  {journal} {Physical Review Letters}\ }\textbf {\bibinfo {volume} {78}},\ \bibinfo {pages} {2275} (\bibinfo {year} {1997})},\ \Eprint {https://arxiv.org/abs/9702027} {arXiv:9702027 [quant-ph]} \BibitemShut {NoStop}%
\bibitem [{\citenamefont {Horodecki}\ \emph {et~al.}(2009)\citenamefont {Horodecki}, \citenamefont {Horodecki}, \citenamefont {Horodecki},\ and\ \citenamefont {Horodecki}}]{Horodecki2009}%
  \BibitemOpen
  \bibfield  {author} {\bibinfo {author} {\bibfnamefont {R.}~\bibnamefont {Horodecki}}, \bibinfo {author} {\bibfnamefont {P.}~\bibnamefont {Horodecki}}, \bibinfo {author} {\bibfnamefont {M.}~\bibnamefont {Horodecki}},\ and\ \bibinfo {author} {\bibfnamefont {K.}~\bibnamefont {Horodecki}},\ }\bibfield  {title} {\bibinfo {title} {{Quantum entanglement}},\ }\href {https://doi.org/10.1103/RevModPhys.81.865} {\bibfield  {journal} {\bibinfo  {journal} {Reviews of Modern Physics}\ }\textbf {\bibinfo {volume} {81}},\ \bibinfo {pages} {865} (\bibinfo {year} {2009})},\ \Eprint {https://arxiv.org/abs/0702225} {arXiv:0702225 [quant-ph]} \BibitemShut {NoStop}%
\bibitem [{\citenamefont {Bekenstein}(1973)}]{Bekenstein1973}%
  \BibitemOpen
  \bibfield  {author} {\bibinfo {author} {\bibfnamefont {J.~D.}\ \bibnamefont {Bekenstein}},\ }\bibfield  {title} {\bibinfo {title} {{Black Holes and Entropy}},\ }\href {https://doi.org/10.1103/PhysRevD.7.2333} {\bibfield  {journal} {\bibinfo  {journal} {Physical Review D}\ }\textbf {\bibinfo {volume} {7}},\ \bibinfo {pages} {2333} (\bibinfo {year} {1973})}\BibitemShut {NoStop}%
\bibitem [{\citenamefont {Belokolos}\ and\ \citenamefont {Teslyk}(2009)}]{Belokolos2009}%
  \BibitemOpen
  \bibfield  {author} {\bibinfo {author} {\bibfnamefont {E.~D.}\ \bibnamefont {Belokolos}}\ and\ \bibinfo {author} {\bibfnamefont {M.~V.}\ \bibnamefont {Teslyk}},\ }\bibfield  {title} {\bibinfo {title} {{Scalar field entanglement entropy of a Schwarzschild black hole from the Schmidt decomposition viewpoint}},\ }\bibfield  {journal} {\bibinfo  {journal} {Classical and Quantum Gravity}\ }\textbf {\bibinfo {volume} {26}},\ \href {https://doi.org/10.1088/0264-9381/26/23/235008} {10.1088/0264-9381/26/23/235008} (\bibinfo {year} {2009})\BibitemShut {NoStop}%
\bibitem [{\citenamefont {Bonnett}(1987)}]{Bennett1987}%
  \BibitemOpen
  \bibfield  {author} {\bibinfo {author} {\bibfnamefont {C.~H.}\ \bibnamefont {Bonnett}},\ }\bibfield  {title} {\bibinfo {title} {Demons, engines and the second law},\ }\href@noop {} {\bibfield  {journal} {\bibinfo  {journal} {Scientific American}\ }\textbf {\bibinfo {volume} {257}},\ \bibinfo {pages} {108} (\bibinfo {year} {1987})}\BibitemShut {NoStop}%
\bibitem [{\citenamefont {Sagawa}\ and\ \citenamefont {Ueda}(2008)}]{Sagawa2008}%
  \BibitemOpen
  \bibfield  {author} {\bibinfo {author} {\bibfnamefont {T.}~\bibnamefont {Sagawa}}\ and\ \bibinfo {author} {\bibfnamefont {M.}~\bibnamefont {Ueda}},\ }\bibfield  {title} {\bibinfo {title} {Second law of thermodynamics with discrete quantum feedback control},\ }\href {https://doi.org/10.1103/PhysRevLett.100.080403} {\bibfield  {journal} {\bibinfo  {journal} {Physical Review Letter}\ }\textbf {\bibinfo {volume} {100}},\ \bibinfo {pages} {080403} (\bibinfo {year} {2008})}\BibitemShut {NoStop}%
\bibitem [{\citenamefont {Holzhey}\ \emph {et~al.}(1994)\citenamefont {Holzhey}, \citenamefont {Larsen},\ and\ \citenamefont {Wilczek}}]{Holzhey1994}%
  \BibitemOpen
  \bibfield  {author} {\bibinfo {author} {\bibfnamefont {C.}~\bibnamefont {Holzhey}}, \bibinfo {author} {\bibfnamefont {F.}~\bibnamefont {Larsen}},\ and\ \bibinfo {author} {\bibfnamefont {F.}~\bibnamefont {Wilczek}},\ }\bibfield  {title} {\bibinfo {title} {{Geometric and renormalized entropy in conformal field theory}},\ }\href {https://doi.org/10.1016/0550-3213(94)90402-2} {\bibfield  {journal} {\bibinfo  {journal} {Nuclear Physics B}\ }\textbf {\bibinfo {volume} {424}},\ \bibinfo {pages} {443} (\bibinfo {year} {1994})},\ \Eprint {https://arxiv.org/abs/9403108} {arXiv:9403108 [hep-th]} \BibitemShut {NoStop}%
\bibitem [{\citenamefont {Dong}\ \emph {et~al.}(2008)\citenamefont {Dong}, \citenamefont {Fradkin}, \citenamefont {Leigh},\ and\ \citenamefont {Nowling}}]{Dong2008}%
  \BibitemOpen
  \bibfield  {author} {\bibinfo {author} {\bibfnamefont {S.}~\bibnamefont {Dong}}, \bibinfo {author} {\bibfnamefont {E.}~\bibnamefont {Fradkin}}, \bibinfo {author} {\bibfnamefont {R.~G.}\ \bibnamefont {Leigh}},\ and\ \bibinfo {author} {\bibfnamefont {S.}~\bibnamefont {Nowling}},\ }\bibfield  {title} {\bibinfo {title} {{Topological entanglement entropy in Chern-Simons theories and quantum Hall fluids}},\ }\href {https://doi.org/10.1088/1126-6708/2008/05/016} {\bibfield  {journal} {\bibinfo  {journal} {Journal of High Energy Physics}\ }\textbf {\bibinfo {volume} {2008}},\ \bibinfo {pages} {016} (\bibinfo {year} {2008})},\ \Eprint {https://arxiv.org/abs/0802.3231} {arXiv:0802.3231} \BibitemShut {NoStop}%
\bibitem [{\citenamefont {Matsueda}(2012)}]{Matsueda2012}%
  \BibitemOpen
  \bibfield  {author} {\bibinfo {author} {\bibfnamefont {H.}~\bibnamefont {Matsueda}},\ }\bibfield  {title} {\bibinfo {title} {{Holographic entanglement entropy in Suzuki-Trotter decomposition of spin systems}},\ }\href {https://doi.org/10.1103/PhysRevE.85.031101} {\bibfield  {journal} {\bibinfo  {journal} {Physical Review E}\ }\textbf {\bibinfo {volume} {85}},\ \bibinfo {pages} {031101} (\bibinfo {year} {2012})}\BibitemShut {NoStop}%
\bibitem [{\citenamefont {Falkovich}\ \emph {et~al.}(2023)\citenamefont {Falkovich}, \citenamefont {Kadish},\ and\ \citenamefont {Vladimirova}}]{Falkovich2023}%
  \BibitemOpen
  \bibfield  {author} {\bibinfo {author} {\bibfnamefont {G.}~\bibnamefont {Falkovich}}, \bibinfo {author} {\bibfnamefont {Y.}~\bibnamefont {Kadish}},\ and\ \bibinfo {author} {\bibfnamefont {N.}~\bibnamefont {Vladimirova}},\ }\bibfield  {title} {\bibinfo {title} {{Multimode correlations and the entropy of turbulence in shell models}},\ }\href {https://doi.org/10.1103/PhysRevE.108.015103} {\bibfield  {journal} {\bibinfo  {journal} {Physical Review E}\ }\textbf {\bibinfo {volume} {108}},\ \bibinfo {pages} {015103} (\bibinfo {year} {2023})},\ \Eprint {https://arxiv.org/abs/2209.05816} {arXiv:2209.05816} \BibitemShut {NoStop}%
\bibitem [{\citenamefont {Tanogami}\ and\ \citenamefont {Araki}(2024)}]{Tanogami2024}%
  \BibitemOpen
  \bibfield  {author} {\bibinfo {author} {\bibfnamefont {T.}~\bibnamefont {Tanogami}}\ and\ \bibinfo {author} {\bibfnamefont {R.}~\bibnamefont {Araki}},\ }\bibfield  {title} {\bibinfo {title} {{Information-thermodynamic bound on information flow in turbulent cascade}},\ }\href {https://doi.org/10.1103/PhysRevResearch.6.013090} {\bibfield  {journal} {\bibinfo  {journal} {Physical Review Research}\ }\textbf {\bibinfo {volume} {6}},\ \bibinfo {pages} {013090} (\bibinfo {year} {2024})},\ \Eprint {https://arxiv.org/abs/2206.11163} {arXiv:2206.11163} \BibitemShut {NoStop}%
\bibitem [{\citenamefont {Kawamori}(2017)}]{Kawamori2017}%
  \BibitemOpen
  \bibfield  {author} {\bibinfo {author} {\bibfnamefont {E.}~\bibnamefont {Kawamori}},\ }\bibfield  {title} {\bibinfo {title} {{Identification of Langmuir wave turbulence-supercontinuum transition by application of von Neumann entropy}},\ }\href {https://doi.org/10.1063/1.4999507} {\bibfield  {journal} {\bibinfo  {journal} {Physics of Plasmas}\ }\textbf {\bibinfo {volume} {24}},\ \bibinfo {pages} {090701} (\bibinfo {year} {2017})}\BibitemShut {NoStop}%
\bibitem [{\citenamefont {Yatomi}\ \emph {et~al.}(2023)\citenamefont {Yatomi}, \citenamefont {Nakata},\ and\ \citenamefont {Sasaki}}]{Yatomi2023}%
  \BibitemOpen
  \bibfield  {author} {\bibinfo {author} {\bibfnamefont {G.}~\bibnamefont {Yatomi}}, \bibinfo {author} {\bibfnamefont {M.}~\bibnamefont {Nakata}},\ and\ \bibinfo {author} {\bibfnamefont {M.}~\bibnamefont {Sasaki}},\ }\bibfield  {title} {\bibinfo {title} {{Data-driven modal analysis of nonlinear quantities in turbulent plasmas using multi-field singular value decomposition}},\ }\href {https://doi.org/10.1088/1361-6587/ace993} {\bibfield  {journal} {\bibinfo  {journal} {Plasma Physics and Controlled Fusion}\ }\textbf {\bibinfo {volume} {65}},\ \bibinfo {pages} {095014} (\bibinfo {year} {2023})}\BibitemShut {NoStop}%
\bibitem [{\citenamefont {KODAHARA}\ \emph {et~al.}(2023)\citenamefont {KODAHARA}, \citenamefont {SASAKI}, \citenamefont {KAWACHI}, \citenamefont {JAJIMA}, \citenamefont {KOBAYASHI}, \citenamefont {YAMADA}, \citenamefont {ARAKAWA},\ and\ \citenamefont {FUJISAWA}}]{Kodahara2023}%
  \BibitemOpen
  \bibfield  {author} {\bibinfo {author} {\bibfnamefont {T.}~\bibnamefont {KODAHARA}}, \bibinfo {author} {\bibfnamefont {M.}~\bibnamefont {SASAKI}}, \bibinfo {author} {\bibfnamefont {Y.}~\bibnamefont {KAWACHI}}, \bibinfo {author} {\bibfnamefont {Y.}~\bibnamefont {JAJIMA}}, \bibinfo {author} {\bibfnamefont {T.}~\bibnamefont {KOBAYASHI}}, \bibinfo {author} {\bibfnamefont {T.}~\bibnamefont {YAMADA}}, \bibinfo {author} {\bibfnamefont {H.}~\bibnamefont {ARAKAWA}},\ and\ \bibinfo {author} {\bibfnamefont {A.}~\bibnamefont {FUJISAWA}},\ }\bibfield  {title} {\bibinfo {title} {{Analysis of Turbulence Driven Particle Transport in PANTA by Using Multi-Field Singular Value Decomposition}},\ }\href {https://doi.org/10.1585/pfr.18.1202036} {\bibfield  {journal} {\bibinfo  {journal} {Plasma and Fusion Research}\ }\textbf {\bibinfo {volume} {18}},\ \bibinfo {pages} {1202036} (\bibinfo {year} {2023})}\BibitemShut {NoStop}%
\bibitem [{\citenamefont {Hasegawa}\ and\ \citenamefont {Wakatani}(1983)}]{Hasegawa1983}%
  \BibitemOpen
  \bibfield  {author} {\bibinfo {author} {\bibfnamefont {A.}~\bibnamefont {Hasegawa}}\ and\ \bibinfo {author} {\bibfnamefont {M.}~\bibnamefont {Wakatani}},\ }\bibfield  {title} {\bibinfo {title} {{Plasma Edge Turbulence}},\ }\href {https://journals.aps.org/prl/abstract/10.1103/PhysRevLett.50.682} {\bibfield  {journal} {\bibinfo  {journal} {Physical Review Letters}\ }\textbf {\bibinfo {volume} {50}},\ \bibinfo {pages} {682} (\bibinfo {year} {1983})}\BibitemShut {NoStop}%
\bibitem [{\citenamefont {Numata}\ \emph {et~al.}(2007)\citenamefont {Numata}, \citenamefont {Ball},\ and\ \citenamefont {Dewar}}]{Numata2007}%
  \BibitemOpen
  \bibfield  {author} {\bibinfo {author} {\bibfnamefont {R.}~\bibnamefont {Numata}}, \bibinfo {author} {\bibfnamefont {R.}~\bibnamefont {Ball}},\ and\ \bibinfo {author} {\bibfnamefont {R.~L.}\ \bibnamefont {Dewar}},\ }\bibfield  {title} {\bibinfo {title} {{Bifurcation in electrostatic resistive drift wave turbulence}},\ }\href {https://doi.org/10.1063/1.2796106} {\bibfield  {journal} {\bibinfo  {journal} {Physics of Plasmas}\ }\textbf {\bibinfo {volume} {14}},\ \bibinfo {pages} {102312} (\bibinfo {year} {2007})},\ \Eprint {https://arxiv.org/abs/0708.4317} {arXiv:0708.4317} \BibitemShut {NoStop}%
\bibitem [{\citenamefont {Nakata}\ \emph {et~al.}(2010)\citenamefont {Nakata}, \citenamefont {Watanabe}, \citenamefont {Sugama},\ and\ \citenamefont {Horton}}]{Nakata2010}%
  \BibitemOpen
  \bibfield  {author} {\bibinfo {author} {\bibfnamefont {M.}~\bibnamefont {Nakata}}, \bibinfo {author} {\bibfnamefont {T.~H.}\ \bibnamefont {Watanabe}}, \bibinfo {author} {\bibfnamefont {H.}~\bibnamefont {Sugama}},\ and\ \bibinfo {author} {\bibfnamefont {W.}~\bibnamefont {Horton}},\ }\bibfield  {title} {\bibinfo {title} {{Formation of coherent vortex streets and transport reduction in electron temperature gradient driven turbulence}},\ }\href {https://doi.org/10.1063/1.3356048} {\bibfield  {journal} {\bibinfo  {journal} {Physics of Plasmas}\ }\textbf {\bibinfo {volume} {17}},\ \bibinfo {pages} {042306} (\bibinfo {year} {2010})}\BibitemShut {NoStop}%
\bibitem [{\citenamefont {Nakata}\ \emph {et~al.}(2011)\citenamefont {Nakata}, \citenamefont {Watanabe}, \citenamefont {Sugama},\ and\ \citenamefont {Horton}}]{Nakata2011}%
  \BibitemOpen
  \bibfield  {author} {\bibinfo {author} {\bibfnamefont {M.}~\bibnamefont {Nakata}}, \bibinfo {author} {\bibfnamefont {T.~H.}\ \bibnamefont {Watanabe}}, \bibinfo {author} {\bibfnamefont {H.}~\bibnamefont {Sugama}},\ and\ \bibinfo {author} {\bibfnamefont {W.}~\bibnamefont {Horton}},\ }\bibfield  {title} {\bibinfo {title} {{Effects of parallel dynamics on vortex structures in electron temperature gradient driven turbulence}},\ }\href {https://doi.org/10.1063/1.3535584} {\bibfield  {journal} {\bibinfo  {journal} {Physics of Plasmas}\ }\textbf {\bibinfo {volume} {18}},\ \bibinfo {pages} {012303} (\bibinfo {year} {2011})}\BibitemShut {NoStop}%
\bibitem [{\citenamefont {Sasaki}\ \emph {et~al.}(2018)\citenamefont {Sasaki}, \citenamefont {Kobayashi}, \citenamefont {Itoh}, \citenamefont {Kasuya}, \citenamefont {Kosuga}, \citenamefont {Fujisawa},\ and\ \citenamefont {Itoh}}]{Sasaki2018}%
  \BibitemOpen
  \bibfield  {author} {\bibinfo {author} {\bibfnamefont {M.}~\bibnamefont {Sasaki}}, \bibinfo {author} {\bibfnamefont {T.}~\bibnamefont {Kobayashi}}, \bibinfo {author} {\bibfnamefont {K.}~\bibnamefont {Itoh}}, \bibinfo {author} {\bibfnamefont {N.}~\bibnamefont {Kasuya}}, \bibinfo {author} {\bibfnamefont {Y.}~\bibnamefont {Kosuga}}, \bibinfo {author} {\bibfnamefont {A.}~\bibnamefont {Fujisawa}},\ and\ \bibinfo {author} {\bibfnamefont {S.-I.}\ \bibnamefont {Itoh}},\ }\bibfield  {title} {\bibinfo {title} {{Spatio-temporal dynamics of turbulence trapped in geodesic acoustic modes}},\ }\href {https://doi.org/10.1063/1.5008541} {\bibfield  {journal} {\bibinfo  {journal} {Physics of Plasmas}\ }\textbf {\bibinfo {volume} {25}},\ \bibinfo {pages} {012316} (\bibinfo {year} {2018})}\BibitemShut {NoStop}%
\bibitem [{\citenamefont {Nakata}\ \emph {et~al.}(2012)\citenamefont {Nakata}, \citenamefont {Watanabe},\ and\ \citenamefont {Sugama}}]{Nakata2012}%
  \BibitemOpen
  \bibfield  {author} {\bibinfo {author} {\bibfnamefont {M.}~\bibnamefont {Nakata}}, \bibinfo {author} {\bibfnamefont {T.-H.}\ \bibnamefont {Watanabe}},\ and\ \bibinfo {author} {\bibfnamefont {H.}~\bibnamefont {Sugama}},\ }\bibfield  {title} {\bibinfo {title} {{Nonlinear entropy transfer via zonal flows in gyrokinetic plasma turbulence}},\ }\href {https://doi.org/10.1063/1.3675855} {\bibfield  {journal} {\bibinfo  {journal} {Physics of Plasmas}\ }\textbf {\bibinfo {volume} {19}},\ \bibinfo {pages} {022303} (\bibinfo {year} {2012})}\BibitemShut {NoStop}%
\end{thebibliography}%

\end{document}